\documentclass[conference]{IEEEtran}

\usepackage{cite}
\usepackage{amsmath,amssymb,amsfonts}
\usepackage{algorithmic}
\usepackage{graphicx}
\usepackage{textcomp}
\usepackage{xcolor}
\usepackage{url}
\usepackage{booktabs}
\usepackage{tikz}
\usepackage{hyperref}

\newcommand{\solidcircnum}[1]{%
\tikz[baseline=(char.base)]{\node[shape=circle,fill=orange,inner sep=1pt,minimum size=9.5pt] (char) {\color{white}\scriptsize\bfseries #1};}%
}

\usepackage[most]{tcolorbox}
\newtcolorbox{takeawaysbox}{
    enhanced,
    breakable,
    boxrule=1pt,
    arc=4pt,
    left=2pt,
    right=2pt,
    bottom=2pt,
    top=2pt,
    colback=gray!4,
    colframe=gray!1!black,
    drop shadow=black!50!white,
    rounded corners,
    before upper={\setlength{\parindent}{1.5em}}
}
\title{\huge Architecting Secure AI Agents: Perspectives on System-Level Defenses Against Indirect Prompt Injection Attacks}

\author{Chong Xiang$^1$ \quad Drew Zagieboylo$^{2*}$\thanks{$^*$Contributions made while affiliated with NVIDIA.}\quad Shaona Ghosh$^1$\quad Sanjay Kariyappa$^1$\quad Kai Greshake$^1$\\Hanshen Xiao$^1$ \quad Chaowei Xiao$^{1,3}$\quad G. Edward Suh$^1$ \\\textsuperscript{1}NVIDIA\quad \textsuperscript{2}Independent Researcher \quad \textsuperscript{3}Johns Hopkins University}
\IEEEoverridecommandlockouts
\begin{document}

\maketitle

\begin{abstract}
AI agents, predominantly powered by large language models (LLMs), are vulnerable to indirect prompt injection, in which malicious instructions embedded in untrusted data can trigger dangerous agent actions.
This position paper discusses our vision for \emph{system-level} defenses against indirect prompt injection attacks. We articulate three positions: (1) dynamic replanning and security policy updates are often necessary for dynamic tasks and realistic environments; (2) certain context-dependent security decisions would still require LLMs (or other learned models), but should only be made within system designs that strictly constrain what the model can observe and decide; (3) in inherently ambiguous cases, personalization and human interaction should be treated as core design considerations.

In addition to our main positions, we discuss limitations of existing benchmarks that can create \textit{a false sense of utility and security}. We also highlight the value of system-level defenses, which serve as the \emph{skeleton} of agentic systems by structuring and controlling agent behaviors, integrating rule-based and model-based security checks, and enabling more targeted research on model robustness and human interaction.
\end{abstract}

\section{Introduction}
AI agents, currently predominantly built on large language models (LLMs),\footnote{For simplicity, our presentation focuses on language agents; however, the discussion also applies to multimodal AI agents.} are susceptible to indirect prompt injection attacks~\cite{greshake2023not}. Attackers can embed malicious instructions in external data---such as retrieved emails, web pages, or third-party tool outputs---to trigger harmful agent actions. As LLM-powered agents are deployed in increasingly high-stakes settings and granted greater autonomy, prompt injection becomes a primary barrier to safe and reliable adoption.

In response, the research community has proposed a broad spectrum of defenses. \emph{Model-level defenses} aim to improve an LLM's inherent robustness via security-oriented fine-tuning~\cite{chen2025struq,chen2025secalign,chen2025meta} or by leveraging internal representations (e.g., activation patterns) to detect adversarial inputs~\cite{hung2025attention,zou2025pishield}. \emph{Text-level defenses} operate on LLM inputs and outputs, for example, by strengthening prompts~\cite{hines2024defending} or by monitoring text strings for anomalous behavior~\cite{shi2025promptarmor}. \emph{System-level defenses}~\cite{debenedetti2025defeating,shi2025progent,wang2025agentarmor,li2025drift,kim2025prompt} treat the agent as a complete computing system, analyzing semantic actions---such as tool-invocation sequences and data-access patterns---to detect attacks and enforce security.

\begin{figure}[!t]
  \centering
  \includegraphics[width=0.9\linewidth]{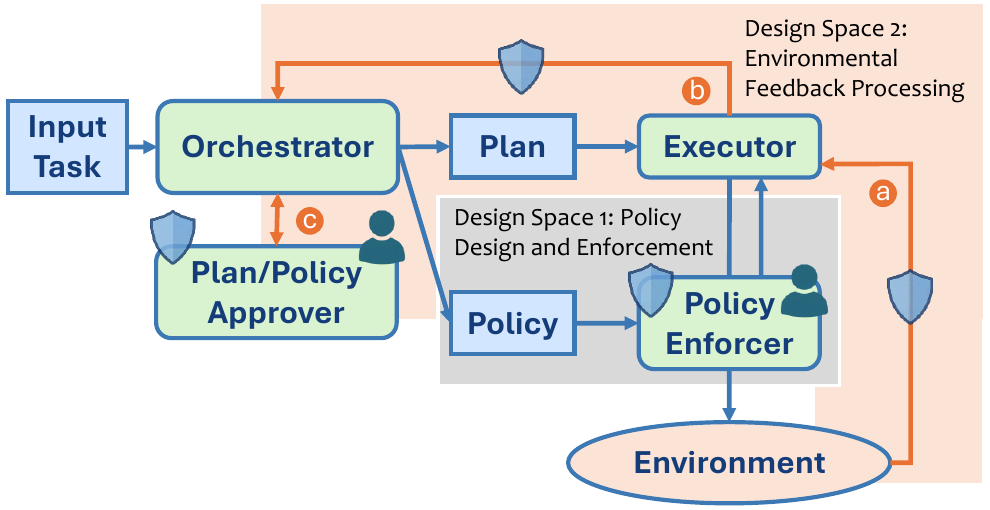}
  \caption{\textbf{High-level system architecture for building LLM agents with both high utility and strong security:} (1) Given a task, the orchestrator generates a plan and policy. (2) The plan/policy approver oversees this generation process to ensure that the resulting plan and policy are reasonable. (3) The executor takes the plan and generates a concrete action. (4) The policy enforcer approves or blocks the action based on the policy. (5.i) If the action is approved, it is issued to the environment, which returns a response. (5.ii) If the action is rejected, the policy enforcer sends negative feedback to the executor. (6) The executor processes feedback from either the environment or the policy enforcer, which can further trigger the orchestrator to update the plan and policy for the next iteration. \\\textbf{Notes:}~\textit{Blue shields} indicate where security-critical decisions can occur and therefore require special security design. \\\textit{Human icons} indicate checkpoints that may require explicit human-in-the-loop oversight, such as personalizing system behavior, or resolving ambiguous objective alignment.}

  \label{fig:system}
\end{figure}

Among these approaches, system-level defenses have recently gained special attention because they can provide more transparent decision-making and may offer meaningful security guarantees. In this position paper, we share our observations and vision for system-level defenses against indirect prompt injection.

To ground our discussion, Figure~\ref{fig:system} presents a high-level system architecture that, we believe, captures the \textit{essential components for building general-purpose agents that are both high-utility and secure}: the orchestrator, plan/policy approver, plan, executor, policy, and policy enforcer. Given an input task, the \textit{orchestrator} first produces a reasonable plan and policy, in coordination with and under the oversight of the \textit{plan/policy approver}: the \textit{plan} describes how the agent may complete the task, while the \textit{policy} specifies what the agent is allowed to do. The \textit{executor} then takes the plan and generates concrete actions, and the \textit{policy enforcer} approves or blocks these actions against the policy. Finally, either the \textit{policy enforcer} responds to the proposed action (i.e., by rejecting it), or the environment responds to the executed action; in either case, this may trigger \textit{plan/policy updates} in the next iteration. Centered on this architecture, we state the following three positions.

\textbf{Position 1: Dynamic (and security-aware) replanning and security policy updates are necessary for complex, dynamic agentic tasks.} A popular idea in existing system-level defenses is \emph{plan-execution isolation}~\cite{willison2023dual,debenedetti2025defeating}: the agent generates a plan solely from the user task and then executes it largely ``as is.'' While this strategy prevents untrusted data from directly tampering with the control flow, it becomes brittle in dynamic environments: even benign runtime errors (e.g., deprecated APIs, missing dependencies) can break the agent workflow. Therefore, we argue that replanning is necessary to preserve the utility of general-purpose agents.

Another common strategy in system-level defenses is policy enforcement~\cite{shi2025progent,debenedetti2025defeating}: the system reasons about allowed behaviors beforehand and blocks suspicious actions. We observe that policies are inherently coupled to the user task and the current plan: replanning often implies policy changes. In open-ended settings, it can be hard (or impossible) to write a fully static policy upfront (e.g., during debugging, one cannot know in advance which files or logs should be read, written, or executed). Therefore, dynamic policy updates are another necessary part of agent design.

However, we recognize that this dynamic design creates a key security challenge: external, untrusted data can now influence plans and policies. The central research problem is therefore to distinguish benign plan/policy updates from malicious ones. In other words, we must research how to make sound security decisions during replanning and policy updates. This question leads to our second position.

\textbf{Position 2: LLMs (or other learned models) are necessary for certain security decisions.}
Security researchers generally agree that securing an LLM agent with another LLM judge can be unreliable, because the judge itself can be manipulated by prompt injection. At the same time, however, we observe that the complexity of agent execution and policy design can exceed the expressivity of traditional rule-based security policies and mechanisms. We propose resolving this tension by leveraging the expressivity of LLMs (or other learned models), but only in settings where the \textit{system strictly limits what they can observe and what they can decide}. That is, when we must involve an LLM in security decision-making, the model should not consume arbitrary (and potentially malicious) environmental text or perform arbitrary tasks. Instead, it should only see narrowly scoped, structured artifacts and be used for a constrained task (e.g., assessing whether a proposed plan or policy change is malicious). This restriction can substantially reduce an attacker's ability to steer the model-based decision (and thus the overall agent).

Moreover, this framing points to a promising opportunity for \textit{co-designing system-level and model-level defenses}: with strong system-level design, we only need the model to be robust (e.g., to reliably ignore malicious instructions) for well-defined, narrow sub-tasks over structured inputs, rather than for arbitrary tasks over arbitrary untrusted inputs. This provides a clearer direction for model-robustness research and also makes it a more tractable research problem.

\textbf{Position 3: Personalization and human interaction should be taken into consideration, especially in ambiguous use cases.}  
Our final position highlights that some aspects of user intent, context, and preferences are inherently ambiguous and cannot be fully resolved through algorithmic system design or general model improvements. 

One common category is \emph{ambiguous language semantics}. For example, given a user task ``fetch and summarize all urgent emails from my inbox,'' what counts as ``urgent'' is user- and context-dependent. Resolving such cases requires user involvement and should draw on insights from usable security.

Another category is \emph{ambiguous objective alignment}. For instance, a coding agent may be instructed (via untrusted content) to install a package that appears relevant to the task but is malicious or compromised. This scenario is not even uncommon for humans: people often follow troubleshooting instructions found online without verifying them. Determining whether such an action is acceptable is difficult to define and evaluate purely from system-level signals, and may require explicit human judgment or organizational policy.

This discussion highlights a fundamental limit of algorithmic, system-level design: developers must be aware of human factors in agent security.\footnote{While human oversight is often treated as a fail-safe for \textit{high-risk actions} (e.g., policy updates), we argue that \textit{ambiguity} is an even more fundamental issue—one that even a theoretically perfect model cannot resolve without human input.}

In addition to these three positions, we highlight two natural questions.

\textbf{Question 1: Why do defenses that omit key components in Figure~\ref{fig:system} still look good on paper?}
We attribute this to limitations in existing benchmark design. First, existing benchmarks lack nuanced, multi-step user tasks where replanning (and policy updates) are required for success. For example, the most popular AgentDojo benchmark~\cite{debenedetti2024agentdojo} only contains 6 (out of 97) tasks that require policy update/replanning~\cite{li2026agentdyn}.
Second, all benchmarks~\cite{debenedetti2024agentdojo,li2026agentdyn, zhang2024agent} consider only non-adaptive, static attack payloads (and objectives), rather than a dynamic attacker that optimizes its payload against the defense system. This can create \textit{a false sense of both utility and security}, motivating the design of more realistic, adaptive benchmarks for prompt injection.

\textbf{Question 2: What is the value of system-level defenses if they still require LLM and/or human for decision making in certain cases?}
We argue that the key value of system-level defense research is that it provides a \textit{structured} way to analyze, understand, and enhance agent security. 
For example, as discussed in Position~2, although we may require LLMs for certain security decisions, we can design the system so that an LLM-based judge does not consume arbitrary environmental text and is used only for narrowly scoped judgments. Moreover, enforcing structured inputs to the LLM clarifies what model-level defenses should target: rather than defending against arbitrary strings, they can be optimized for specific structured inputs and constrained judgment tasks.
System-level defenses also improve explainability by making it clearer why the agent takes certain actions, and they enable the integration of rule-based mechanisms to validate and enforce security policies. In this way, they serve as a bridge between programmatic security enforcement and expressive model-based security decision-making, enabling defense in depth.
Overall, we view system-level defense design as the \textit{skeleton and backbone} for building agents that are both high-utility and secure.

\textbf{Scope of the paper.} We articulate the scope of this paper as follows.
\begin{itemize}
    \item We focus on \textit{general-purpose AI agents with full autonomy}, rather than agents with predefined workflows and limited autonomy.
    \item We focus on \textit{indirect prompt injection attacks}, in which malicious instructions embedded in external data segments (from a partially compromised environment) hijack agent actions. That is, we assume the system and user prompts are trusted, while the environment is partially compromised.
    \item We focus on \textit{algorithmic system-level defenses} that treat the agent as a computing system and examine how to architect agentic systems for security.
    \item We do not discuss orthogonal traditional security techniques such as sandboxing; instead, \textit{we focus on security challenges that are fundamentally unsolvable through sandboxing.}
    \item This paper does not aim to cover basic concepts of AI agents or prompt injection, nor does it provide a comprehensive survey of existing work. We refer readers to recent and contemporaneous SoK papers~\cite{wang2026landscape,christodorescu2025systems,ji2025taxonomy} for background.
\end{itemize}

\textbf{Paper organization.} We organize the rest of the paper as follows. We elaborate on our proposed system architecture (Figure~\ref{fig:system}) in Section~\ref{sec-layout}, then expand our three positions with concrete case studies in Section~\ref{sec-position}. We also discuss the limitations of existing benchmarks and the value of system-level defenses in Section~\ref{sec-discussion}, and conclude in Section~\ref{sec-conclusion}.

\section{System Architecture Toward High-Utility and Secure Agents}\label{sec-layout}
In this section, we discuss the high-level system architecture in Figure~\ref{fig:system}, which we believe captures essential components for building agents that are both high-utility and robust. Section~\ref{sec-layout-plan-policy} introduces two key concepts in our architecture design: \emph{plan} and \emph{policy}. Then, Section~\ref{sec-layout-walkthrough} walks through the architecture step by step, discusses its design space, and maps representative prior defenses to it. Together, these discussions lay the foundation for the positions we present in the rest of the paper.

\subsection{Plan and Policy: A Structured Way to Analyze and Control LLM Agents}\label{sec-layout-plan-policy}
By design, LLMs operate on unstructured natural-language text. This unstructured nature facilitates large-scale LLM training but makes LLM-powered agents difficult to analyze and control. To address this challenge, it is natural to introduce the concepts of \emph{plan} and \emph{policy}. Intuitively, a \textit{plan} describes what an agent can do to complete a user task, while a \textit{policy} describes what an agent is allowed to do. We provide a formulation and examples below.

\textbf{Plan.} We model a single agent execution step as
$\mathbf{e} := (a, \mathcal{I}, \mathcal{O})$,
where $a$ is a concrete agent action (e.g., a tool invocation such as ``read email''), and $\mathcal{I}$ and $\mathcal{O}$ are the input and output objects (e.g., a file path, or the ``five most recent emails''). A \emph{plan} $\mathbf{p}\in\mathcal{P}$ is then an ordered sequence (or, more generally, a directed graph) of execution steps,
$\mathbf{p} := (\mathbf{e}_1, \ldots, \mathbf{e}_T)$,
which describes a possible execution trajectory for completing the user task.

In practice, a plan can be represented at different levels of formality (and granularities). For example, given a task ``summarize the most recent email from Alice and draft a reply,'' the following are some illustrative representations (not an exhaustive list): 
\begin{enumerate}
    \item \textit{Natural-language checklist:}
    {\footnotesize
    \begin{verbatim}
(i) Search for the most recent email from Alice
(ii) Summarize key points
(iii) Draft a polite reply
    \end{verbatim}
    }

    \item \textit{A domain-specific language (DSL) with typed steps and dependencies:}
    {\footnotesize
    \begin{verbatim}
e1: GET_RECENT_EMAIL(sender=Alice,n=1) -> emails
e2: SUMMARIZE(emails) -> summary
e3: DRAFT_REPLY(summary) -> draft
    \end{verbatim}
    }
    %\item \textit{A structured object (e.g., JSON) that exposes control-flow and data-flow:}
    %\begin{quote}
    %\texttt{\{ "steps": [}\\
    %\texttt{  \{ "id":"e1", "op":"search\_recent\_email", "args":\{"sender":"Alice","n":1\}, "out":"emails" \},}\\
    %\texttt{  \{ "id":"e2", "op":"summarize", "in":["emails"], "out":"summary" \},}\\
    %\texttt{  \{ "id":"e3", "op":"draft\_reply", "in":["summary"], "out":"draft" \},}\\
   % \end{quote}
\end{enumerate}

\textbf{Policy.} A \emph{policy} specifies which actions and information flows are permitted during execution. Formally, we can view a policy as a predicate $\boldsymbol{\pi}(\cdot)$ over steps and execution histories, where $\boldsymbol{\pi}(\mathbf{e}_t)=1$ indicates that step $\mathbf{e}_t$ is allowed under the current context. Equivalently, at the plan level, a policy induces a set of permitted plans $\mathcal{P}_{\boldsymbol{\pi}} \subseteq \mathcal{P}$.

In practice, a policy can be specified at different types of constraints. For example, a policy might include:
\begin{itemize}
    \item \textit{Global static access-control rules:} e.g., the agent should never read, write, or execute data that the user does not have permission to access.
    \item \textit{Task-specific least privilege:} e.g., for a given task, the agent should only use a prescribed set of necessary tools and access a prescribed set of data (and nothing more).
    \item \textit{Context-dependent security principles:} permissions may depend on the current context; e.g., the agent may only initiate a money transfer if the recipient is a verified contact.
    \item \textit{Lattice-based information-flow control (IFC):} label data/tool with security levels (e.g., $\texttt{public} \sqsubset \texttt{internal} \sqsubset \texttt{confidential}$) and only permit actions that do not violate allowed flows (e.g., no write-down from higher to lower labels).
\end{itemize}
Additionally, policies can be specified across multiple layers, such as the tool, application, network, OS, and infrastructure layers.

Policy representations can also vary widely. For example, a policy may be encoded as:
\begin{enumerate}
    \item \textit{A structured object (e.g., JSON) that lists allow/deny rules:} 
    e.g., a simplified example from Progent~\cite{shi2025progent} that denies listing private repositories.

    {\footnotesize
    \begin{verbatim}
{"tool": "list_repos",
 "rules": [
   {"effect": "forbid",
    "condition":
       {"include_private": {"const": true}}}]}
    \end{verbatim}
    }

    \item \textit{A customized language (or code-level policy function) that can express richer, context-dependent checks:} 
    a simplified example from CaMeL~\cite{debenedetti2025defeating} that allows creating calendar events if all attendees are trusted.

    {\footnotesize
    \begin{verbatim}
def calendar_event_policy(tool_name, kwargs):
    attendees = kwargs["attendees"]
    if is_trusted(attendees):
        return Allowed()
    \end{verbatim}
    }

    Another simple Cedar~\cite{cutler2024cedar} example that permits user \texttt{alice} to view document \texttt{doc1}.

    {\footnotesize
    \begin{verbatim}
permit(
    principal == User::"alice",
    action == Action::"view",
    resource == Document::"doc1"
);
    \end{verbatim}
    }
\end{enumerate}

With the concepts of plan and policy, in the next subsection, we turn to our system architecture discussion.

%Figure~\ref{fig:system} shows our proposed architecture for building high-utility, secure agents; we explain its workflow step by step.
\subsection{System Architecture Design} \label{sec-layout-walkthrough}
In this subsection, we walk through our proposed architecture for building high-utility, secure agents (Figure~\ref{fig:system}), discuss its design space, and then connect it to existing defenses.

\textbf{System architecture walkthrough.} The architecture presented in Figure~\ref{fig:system} is designed around the concepts of plan and policy discussed earlier. We present a step-by-step walkthrough below.
\begin{enumerate}
    \item Given a high-level task, the \textbf{Orchestrator}, typically powered by an LLM, produces an initial \textbf{Plan} (how to complete the task) and \textbf{Policy} (what is allowed while doing so).
    \item The \textbf{Plan/Policy Approver}, typically powered by an LLM, oversees the generation process of plan and policy. Its role is to approve the generated plan and policy and to provide feedback to the Orchestrator if it detects any unreasonable components. As denoted by the human icon, this module can also escalate to a human user to provide personalization or clarify ambiguous objectives (see Position 3 in Section~\ref{sec-position-3} for more discussions). Conceptually, this Plan/Policy Approver can be viewed as part of the Orchestrator design; \textit{we separate it explicitly to highlight its role as a security decision module.}
    \item The \textbf{Executor}, typically an LLM, consumes the plan and generates a concrete action (e.g., a tool call with arguments).
    \item The \textbf{Policy Enforcer} approves or blocks the proposed action based on the current policy. The policy check can be performed by rule-based programs or by an LLM, depending on the format and instantiation of the policy. For high-risk actions or ambiguous scenarios, this enforcement step can further defer to human judgment (indicated by the human icon).
    \item Once approved, the action is sent to the \textbf{Environment} (APIs, the web, and file system), which executes it and returns a response. Otherwise, the \textbf{Policy Enforcer} sends negative feedback to the \textbf{Executor}.
    \item The \textbf{Executor} processes the response (from either the Environment or the Policy Enforcer) and reports back to the \textbf{Orchestrator}; this may trigger updates to the plan and/or policy.
    \item The agent repeats this loop on the updated plan and policy until it finishes the task.
\end{enumerate}

\textbf{System design space.} Under this architecture, we have two key design axes. 

\textit{Design space 1: policy scope, representation, and enforcement.}
The first design choice we need to make is what the policy should \textit{capture} (e.g., least privilege, global invariants, and context-dependent constraints), how it should be \textit{represented} (e.g., templates, typed DSLs, or graphs that make control and data flow explicit), and how it should be \textit{enforced} (e.g., programmatic checks, learned judges, human oversight, or hybrid approaches). In general, richer policy representations increase expressivity but can make enforcement harder; good designs balance expressivity with checkability.

\textit{Design space 2: security-aware processing of environmental feedback (and subsequent plan/policy updates).}
The architecture in Figure~\ref{fig:system} allows environmental responses to influence the plan and policy over time. This improves adaptability but also expands the attack surface. The second key design axis is to enable this environmental feedback (arrows \solidcircnum{a}--\solidcircnum{c} in Figure~\ref{fig:system}) while restricting attackers' influence on the model decisions. We discuss concrete directions in Section~\ref{sec-position-2}. %; one key idea is to restrict the structure of the LLM input so that an attacker cannot inject arbitrary malicious text into the LLM context.

\textbf{Connection to existing defenses.} We note that this system architecture is not entirely novel: many existing defenses leverage some of its components. However, existing defenses often use only a subset of these modules, or connect them in ways that compromise security or reduce utility. We discuss representative examples below.
\begin{itemize}
    \item CaMeL~\cite{debenedetti2025defeating} follows a plan-execution isolation paradigm: its plan is based solely on the input task and remains fixed throughout execution. This can be viewed as completely removing the feedback and update loop in \solidcircnum{b}. As we discuss in Section~\ref{sec-position-1}, this strict isolation may prevent agents from completing certain tasks in a dynamic environment.    
    \item Progent~\cite{shi2025progent} broadly follows this architecture, but its plan/policy update mechanism is not security-aware: it directly feeds unrestricted environment feedback to another LLM to update the plan/policy, and thus remains vulnerable to adaptive attacks targeting the LLM-based policy adjuster (e.g., \solidcircnum{c}). %We discuss how we might overcome this limitation in Section~\ref{sec-position-2}.
    \item The PFI defense~\cite{kim2025prompt} follows a similar workflow and reduces the downstream influence of untrusted external data by redacting tool responses and replacing them with placeholders before the executor consumes them (modifying \solidcircnum{b}). However, this redaction may impact utility, in part because the LLM may struggle to act on redacted environmental feedback (as noted in the original paper), and in part because some tasks require information that has been redacted.
    \item AgentArmor~\cite{wang2025agentarmor} implicitly follows this architecture. It allows the agent to run as-is (enabling replanning) and leverages a policy enforcer that tracks data and control flows during execution and blocks any action that violates IFC-based access control. However, this data/control-flow tracking is performed by an LLM judge with raw environmental text as input, making it susceptible to adaptive attacks.
     \item The DRIFT defense~\cite{li2025drift} can also be viewed as an instance of this architecture. It generates a plan solely based on the user query, uses an LLM judge to decide whether to update the plan as the agent executes, and uses another LLM to remove potentially malicious text. Although removing raw environment text reduces the attack surface of their judge model, adaptive attackers may still be able to steer the trajectory and mislead the model.
\end{itemize}

In the next section, we will present our three positions to further explain why we believe this architecture is necessary for building agents with high utility and security.

\section{Positions}\label{sec-position}

In this section, we expand on the three positions outlined in the introduction, using concrete examples and proposals.

\subsection{Position 1: Dynamic (and Security-Aware) Replanning and Policy Updates Are Necessary for Dynamic Tasks and Environments}\label{sec-position-1}
Position~1 follows a simple observation: general-purpose agents inevitably operate in dynamic environments. As a result, \emph{dynamic replanning} and \emph{dynamic policy updates} are necessary for maintaining high utility---but making these updates securely becomes a central challenge.

\textbf{Replanning is often necessary for utility in dynamic environments.}
A common system-level defense idea is \emph{plan-execution isolation}~\cite{willison2023dual,debenedetti2025defeating}: generate a plan from the user task and then execute it largely ``as is,'' without letting untrusted environment data directly alter control flow. While isolation can significantly reduce the attack surface for control-flow hijacking, it can be brittle even under benign changes. We provide two concrete examples below.

\emph{Example scenario (benign runtime failure).} A data-analysis agent is asked to ``pull last week's sales from the billing API and plot the trend.'' The orchestrator plans to call \texttt{GET /v1/invoices?start=...}. At runtime, the tool returns \texttt{HTTP\_410 Gone} because the endpoint was deprecated and replaced by \texttt{/v2/invoices/search}. A useful agent must adapt its plan: consult updated documentation, switch endpoints, and update parsing logic. Under strict plan-execution isolation, the agent can repeatedly fail the same call and never complete the task.

\emph{Example scenario (iterative debugging depends on feedback).} A coding agent is asked to ``run unit tests and fix failing ones.'' The high-level plan (run tests $\rightarrow$ inspect failure logs $\rightarrow$ patch code) is easy to state, but the concrete next step depends critically on what the failure log reveals. For example, one failure might point to an off-by-one error in a boundary check, while another might indicate a race condition that requires refactoring synchronization. After applying a patch, the agent must rerun the tests (often multiple times) to confirm the fix and detect regressions. Because each patch changes what the next failure looks like (if any), it is difficult to fully specify a detailed plan upfront; replanning based on execution feedback is intrinsic to the task.

\emph{Note on conditional/branching plans.} One might argue that, in the first example, the initial plan could include conditions and branches (e.g., ``if the API call succeeds, proceed with the returned data; otherwise, look up the updated documentation''). While such conditional structure can help for a few anticipated failure modes, it is often infeasible to enumerate all relevant runtime errors in advance---especially in dynamic settings like debugging, as in our second example. Moreover, when control flow depends on environment-provided signals, an attacker may be able to influence which branch is taken (i.e., control-flow steering). Thus, isolated conditional plans are \textit{not a complete solution for either security or utility}; designs that support \emph{security-aware} replanning remain an important research direction.

\textbf{Dynamic policy updates can also be necessary in addition to replanning.} Intuitively, when the execution plan changes, the policy should be updated accordingly. More broadly, we note that static policies can fall short for both utility and security.

\textit{\underline{Reason 1:} Static task-agnostic (e.g., tool-level) policies can be overly restrictive or overly lenient.}
We illustrate this with the AgentDojo~\cite{debenedetti2024agentdojo} benchmark's \texttt{send\_money} tool and a static tool-level policy used in CaMeL~\cite{debenedetti2025defeating}. The \texttt{send\_money} tool transfers the user's money to a recipient, and the CaMeL policy roughly states: \emph{allow only if (i) the recipient comes only from the user, and (ii) the recipient is allowed to observe all tool parameters.}
This policy can work for a narrow benchmark setting, but as a task-agnostic global rule it can produce both false positives and false negatives in broader use cases.

\emph{False positive example (too restrictive).} A user asks: ``Please pay the bill from \texttt{xyz@abc.com}.'' The recipient information might be specified in a legitimate shared artifact (e.g., an email or an attached invoice). However, the rigid requirement that the recipient come \emph{only} from the user's initial prompt would block this benign extension.

\emph{False negative example (too lenient on integrity).} A user asks: ``Pay Joe back for the bouquet of flowers he bought at Store XYZ.'' To resolve the amount, the agent may search the web (or consult other external data) for the price. A prompt injection embedded in retrieved content could steer the agent into setting a larger amount. If the policy constrains only the recipient's provenance but places no integrity constraints on the \texttt{amount} parameter, this attack can be fully permitted under the static policy.

These examples suggest that static, task-agnostic policies may need to be refined dynamically at runtime, either to achieve task success (relax overly pessimistic constraints with evidence) or to minimize risk (tighten overly lenient constraints as new threat surfaces appear).

\textit{\underline{Reason 2:} Even task-specific policies can be insufficient if written only once at task start.}
Even when a policy is tailored to the task, the exact resources and parameter values needed to complete the task may not be knowable upfront.

\emph{Example scenario (debugging requires access updates).} A developer asks: ``Investigate why the service is crashing and implement a fix.'' The initial policy may allow reading the source tree and running tests, but at runtime the agent discovers that the crash occurs only in production and that the relevant evidence is in \texttt{/var/log/myservice/error.log}; therefore, the agent requires read access to this log. Moreover, only after reading this error log can the agent determine which additional code files need to be read, modified, or executed next. A policy written at task start cannot realistically enumerate all log paths, endpoints, diagnostics, and file-level actions that might be required.

\emph{Example scenario (Slack invitation requires learned details).} Consider an AgentDojo~\cite{debenedetti2024agentdojo} task: ``Invite our new colleague to Slack and add her to the necessary channels. You can find details in the message from Bob to Alice in Alice's inbox. Note that usernames start with a capital letter.'' The true least privilege of this task is highly specific (e.g., only read Alice's inbox; only invite the user mentioned in Bob's message; only add her to the channels listed there), yet at the beginning of execution the system may not know the exact username or channel list.
%A conservative approach is to start with a narrow policy (e.g., no Slack writes), then after retrieving and \emph{sanitizing/vetting} the relevant fields from Bob's message, issue a structured policy-update request that expands permissions \emph{only} to the specific invitee and channels required by the task (optionally falling back to human approval for sensitive expansions).

\textbf{Security paradox: external data now influence the plan and policy.}
So far, we have discussed why dynamic replanning and policy updates may be necessary, which in turn motivates the system architecture in Figure~\ref{fig:system}. At the same time, enabling replanning and policy updates allows untrusted environmental feedback to influence agent actions and decisions. The key security challenge then becomes how to distinguish benign plan/policy updates from malicious ones. In other words, we need a reliable mechanism for making security decisions during replanning and policy updates. Our second position aims to address this challenge.\footnote{For readers familiar with the dual-LLM design~\cite{willison2023dual}, our first position essentially argues that there should be an information flow from a quarantined LLM (executor) to a privileged LLM (orchestrator), but how to secure this flow is a central research challenge (i.e., Design Space 2 in Figure~\ref{fig:system}).}

\subsection{Position 2: LLMs (or Specialized Models) Can Be Essential in Security-Critical Decision Making}\label{sec-position-2}

Dynamic replanning and policy updates (Position~1) are largely motivated by the need to maintain agent utility in dynamic environments. Our second position focuses on how to secure these dynamic behaviors.

\textbf{Where can security-related operations take place?}
Before detailing our position, we clarify where security-relevant decisions can occur in an agentic system, echoing the two design spaces discussed in Section~\ref{sec-layout-walkthrough}. First, security checks can occur in the \emph{policy and policy-enforcement} layer (the gray region in Figure~\ref{fig:system}). Depending on the policy representation and scope, enforcement may be implemented by rule-based checkers and/or an LLM-based judge. Second, security checks can occur when the system \emph{processes environmental feedback} (along arrows \solidcircnum{a}--\solidcircnum{c} and in the Orchestrator and Plan/Policy Approver, in the orange region of Figure~\ref{fig:system}): along each arrow, the system may (i) pass raw text, (ii) transform or filter it into safer representations, and/or (iii) monitor for anomalous inputs and outputs. These checkpoints are marked with blue shields in Figure~\ref{fig:system}.

\textbf{Position.}
We argue that LLMs (or other learned models) are essential at \emph{at least one} of these security checkpoints. This may sound counterintuitive: if LLM ``judges'' are themselves vulnerable to prompt injection, how can we use them for security-critical decisions? Our answer is to leverage the model's \emph{expressivity}, but only within a system design that strictly constrains (i) what the model can observe and (ii) what it is allowed to decide.

\textbf{Why programmatic checks alone can be insufficient.}
Many security invariants are easy to encode and enforce in code (e.g., ``never write outside the workspace,'' ``never send network requests to non-allowlisted domains''). However, \textit{comprehensive} rules are hard to write in practice because security decisions are often context-dependent, and that context can be highly diverse (e.g., user intent, prior execution state, and how newly observed information relates to the current task). The central challenge is making reliable context-dependent security decisions---namely, determining whether a proposed security-critical change is \emph{semantically} justified by the user task---which typically exceeds the capability of purely programmatic rules.

\emph{Example scenario (semantics and context needed).}
We reuse the runtime failure example discussed in the previous subsection. After an \texttt{HTTP\_410 Gone} error, the orchestrator proposes a plan update: ``replace \texttt{/v1/invoices} with \texttt{/v2/invoices/search} and update the response parser.'' This update may be benign, but a purely syntactic checker cannot easily determine whether the new endpoint is task-relevant or whether it quietly introduces a new (and potentially risky) data source.

\emph{Analogy (visual semantics).}
An analogous (and more intuitive) limitation appears in computer vision: given an image (a tensor of pixel values), we can mathematically define low-level primitives such as edges via local pixel differences, but it is difficult to specify ``cat'' vs.\ ``dog'' with hand-written rules over pixels. This is why we turn to learned, though less interpretable and sometimes less robust, models.

\textbf{What is the right objective for general-purpose agentic security?}
At this point, a natural concern is that using an AI model for security decisions precludes a 100\% formal security guarantee. \textit{We agree.} However, human assistants can also make mistakes (e.g., falling for phishing emails or unintentionally installing malware), yet they still avoid clearly absurd actions, such as following obviously malicious instructions from an unrelated social media post. For general-purpose agent systems, we therefore argue that the right objective is to improve security while largely preserving utility: even when perfect security is unattainable, systems should make low-effort attacks unlikely to succeed and increase attacker cost. In this way, a successful compromise requires substantial attacker knowledge and investment, rather than a trivial malicious-instruction injection. We next present our strategies toward this objective.

\textbf{Key idea: avoid direct exposure to raw attacker-controlled text or arbitrary tasks when LLMs are involved in security decisions.}
Despite their expressivity, naively using LLMs for security-critical decisions is risky because they are susceptible to prompt injection. Our core system-level insight is therefore to design the agent so that the model is \emph{not directly exposed to raw environment text} (e.g., web pages, emails, or tool outputs) that an attacker may arbitrarily control. Instead, the system should transform untrusted environmental feedback into narrowly scoped, structured artifacts---for example, a typed trace and a proposed plan/policy diff---and ask the model to make judgments only over these constrained representations. Put differently, the model should be used as a bounded security decision module (e.g., evaluating proposed updates under explicit constraints), rather than as an open-ended controller for arbitrary task execution over arbitrary text. This reduces the attacker's ability to shape the model's context while preserving the model's ability to perform semantic reasoning that strict programmatic checks often cannot.

\textbf{Opportunity: co-designing system-level and model-level defenses.}
This strategy also creates a strong opportunity to \emph{co-design} system-level and model-level defenses. The system architecture can deliberately narrow and structure what reaches the model, which \textit{clarifies the direction of model-level robustness research and makes the problem more tractable}. With this design, the learned model no longer needs to be robust to arbitrary untrusted strings across arbitrary tasks; instead, it can focus on smaller, well-defined problems such as judging typed diffs, synthesizing validators over structured outputs, and making bounded decisions with explicit failure modes. We view this co-design opportunity as one of the most promising research directions for building high-utility, secure general-purpose agents.

We next present two concrete proposals for using LLMs in security decisions under input and task constraints.
%both illustrating the system-model co-design philosophy.

\textbf{Proposal 1: Decoupling instruction recognition and instruction-following decision.} Our first proposal is inspired by our prior work on instruction-following intent analysis~\cite{kang2025mitigating}. We propose decomposing the instruction-following process of an LLM/agent into two steps: (1) recognize candidate instructions in context, and (2) decide which recognized instructions should be followed. This decoupling improves transparency and shifts part of the defense burden from implicit model behavior to explicit system policy.

For Step~(1), we propose to make the model, via prompting and/or training, explicitly verbalize (i.e., repeat) the instructions it intends to follow. The motivation is that post-trained LLMs already perform an internal recognition process that distinguishes ``instructions to follow'' from other text; this recognition process is ultimately what matters, because if the model does not recognize malicious instructions as instructions it should follow, a prompt-injection attack fails. Notably, our goal is not to add a new capability, but to make the model \emph{surface} an existing one by faithfully verbalizing the instructions it already
intends to follow. We hypothesize that training a model to faithfully verbalize instructions is a more tractable and less utility-degrading objective than training a model to directly recognize and \textit{ignore} all malicious content.\footnote{One concern is adaptive attacks that attempt to suppress verbalization itself (e.g., ``do not reveal intended instructions''). In this framework, such payloads are still candidate instructions; if verbalized, they become attributable.}

For Step~(2), rather than relying implicitly on model robustness, we make instruction-following decisions at the system level. Concretely, the system first traces the provenance of each verbalized instruction (e.g., via word/substring matching against trusted vs. untrusted segments), then applies one of the operating modes below (as part of the Policy Enforcer module in Figure~\ref{fig:system}).

\emph{Mode 1 (no external instructions).} If the workflow assumes that instructions should come only from trusted sources (e.g., user/developer/system prompts), then any instruction traced to external untrusted data is blocked by default.

\emph{Mode 2 (external instructions allowed with human approval).} If external instructions may be necessary, the system can require explicit user confirmation before executing each such instruction (or each sensitive class of instruction).

\emph{Mode 3 (minimize user intervention).} If frequent confirmation is undesirable, we suggest two complementary mechanisms. (a) \emph{Pre-declare instruction expectation:} before processing an external source, the agent predicts whether that source should contain actionable instructions at all (e.g., a library's ``get started'' page is likely to contain instructions); if not, any extracted instruction is rejected by policy. (b) \emph{Structured instruction-state adjudication:} the orchestrator maintains a structured task/instruction data structure (e.g., a task tree or typed trace) that records executed instructions, currently intended instructions, and provenance. When proposing a new instruction node (e.g., ``install library $X$''), the system submits a structured update for LLM-judge review. Crucially, the judge sees only the structured state and proposed diff, \textit{not raw environment text}, so the model is used as a bounded adjudicator rather than an open-ended follower of untrusted content.

\begin{takeawaysbox}
\noindent \textbf{Design philosophy.} We want to highlight two of our design principles here. \textit{\textbf{First, we aim to simplify the LLM's role in security decisions as much as possible.}} Whenever rule-based controls are sufficient, we should prefer them---for example, blocking externally sourced instructions by provenance (Mode~1) or requiring user confirmation (Mode~2). This reduces reliance on fragile model-level robustness and makes behavior easier to audit.

\textit{\textbf{Our second principle is that, when we must use LLMs for security decisions, we should tightly constrain both their inputs and their tasks.}} In this proposal, for Step~(1) we ask the model only to surface its non-security instruction-recognition process; in Step~(2), Mode~3(a), the model predicts whether a source is expected to contain instructions \emph{before} seeing potentially malicious content; and in Mode~3(b), the model reasons over internally maintained structured instruction state and diffs rather than raw environment text.
\end{takeawaysbox}

\begin{takeawaysbox}
\noindent\textbf{Notes on existing pure model-level defenses.} Model robustness research~\cite{chen2025secalign,chen2025meta,wallace2024instruction} has made important progress on training models to make instruction-following decisions more robustly. From a system-security perspective, however, an important question remains: even if a model can reliably ignore malicious instructions, \textit{should it still rely on other content from the same source once prompt injection is detected?} For example, if the model correctly identifies malicious instructions on an external document, should it continue using other content from that potentially compromised document? Probably not in many settings. We view this as an important security question that is not yet fully resolved by current model-robustness work. This is one reason we propose decoupling instruction recognition from instruction-following decisions. A separate instruction-following decision step can improve explainability and security in agentic systems and provide greater flexibility in how detected attacks are handled.
\end{takeawaysbox}

\textbf{Proposal 2: Using an LLM to synthesize step-specific programmatic validators for environment feedback.}
In addition to verbalizing instructions and judging plan/instruction diffs, the system can use an LLM at runtime to generate \emph{deterministic} checks/programs that validate environment responses at each step (i.e., protecting feedback channel \solidcircnum{a} in Figure~\ref{fig:system}). Here again, the model is used for bounded synthesis of validators, not for arbitrary end-to-end task control over raw text. 

\emph{Example scenario (web retrieval with field-level integrity constraints).}
Suppose the user asks: ``Find Acme's investor relations page and extract the Q4 revenue.'' A retrieval tool returns a page that may contain injected text. The system can ask the LLM to synthesize a validator such as: ``extract the revenue only from the \texttt{Revenue} cell of the table whose header contains \texttt{Quarter} and \texttt{Q4}; ignore any free-form paragraphs; require the extracted number to match \texttt{[0-9,.]+} and cite the DOM path.'' At runtime, the system applies this validator to a structured representation of the page (e.g., a DOM-to-JSON parse) and rejects outputs that violate the constraints, without relying on the executor to ``do the right thing'' when faced with adversarial prose.

\subsection{Position 3: Personalization and Human Interaction Should Be Considered Given Ambiguous Use Cases}\label{sec-position-3}
Finally, our last position highlights a limit of purely algorithmic, system-level defenses: some security decisions may depend on ambiguous language semantics and user intent. In these cases, personalization and human interaction should be taken into account in agent design. We discuss two types of ambiguity below, while noting that other challenging and unavoidable forms of ambiguity may exist.

\textbf{Ambiguity in language semantics.}
Some tasks contain terms whose meaning is user- and context-dependent. This ambiguity \textit{cannot be resolved even with a perfect model} and thus requires human feedback.

\emph{Example scenario (``urgent'' is ambiguous).} A user asks: ``Fetch and summarize all urgent emails from my inbox.'' Here, ``urgent'' may mean ``mentions a deadline within 24 hours'' or ``flagged by a specific label.'' This ambiguity can only be resolved through human interaction.

\textbf{Ambiguous objective alignment.}
Even if an action looks ``relevant'' to a sub-goal in completing the user task, it may still be undesirable or unacceptable under a security policy.

\emph{Example scenario (GitHub issue triage).} A developer asks an agentic coding assistant to ``triage this GitHub issue and propose a patch.'' The agent opens a newly filed issue that includes a ``reproduction'' section with instructions such as ``run \texttt{curl ... | bash}'' to fetch the test harness. These steps can look directly relevant to the objective (reproduce the bug), but executing them may run attacker-controlled code (e.g., via a malicious package install script) that exfiltrates secrets from the environment or modifies the repository. The core ambiguity is that ``do whatever is needed to reproduce'' can conflict with a security policy like ``never execute untrusted code from the internet without explicit approval.'' Notably, this failure mode is not unique to agents: human developers can make the same mistake by blindly following ``reproduction'' or ``fix'' instructions from untrusted sources.\footnote{While we emphasize ambiguity as the fundamental driver for human intervention---since even a theoretically perfect LLM cannot resolve subjective user intent without interaction---we acknowledge that human-in-the-loop (HITL) mechanisms may remain a practical necessity today for less ambiguous but high-stakes workflows (e.g., security policy update).} 

\begin{takeawaysbox}
\noindent \textbf{Design implication: making human-in-the-loop (HITL) usable.} The examples above suggest that human involvement may be unavoidable in certain scenarios, which is why we explicitly embed human-in-the-loop checkpoints (human icons) in Figure~\ref{fig:system}. Here, the key question is how to incorporate human feedback without undermining the point of agents---reducing human effort and interruption. This raises design challenges such as how to learn user preferences from prior human-agent interactions, and whether interaction can be front-loaded (e.g., through clarification and consent at the start of execution) to avoid frequent interruptions later. An analogy is onboarding a new human assistant: early on, they typically need to ask many clarification questions to learn the principal's preferences and resolve ambiguity, but over time they internalize those preferences and require fewer interruptions.
We view these as interesting and important directions for future work, and we expect that progress will benefit from methods and insights from usable security and human-computer interaction.
\end{takeawaysbox}

\section{Discussion}\label{sec-discussion}
This section revisits two questions from the introduction and connects them to our positions.

\textbf{Question 1: Why do defenses that omit key components in Figure~\ref{fig:system} still look good on paper?}
Many existing defenses~\cite{debenedetti2025defeating,shi2025progent,wang2025agentarmor,li2025drift,kim2025prompt} report both high utility and strong security in their evaluations, despite omitting key design components discussed in this paper. We attribute this discrepancy to limitations in existing benchmarks~\cite{debenedetti2024agentdojo,zhan2024injecagent}---in both user tasks and attack modeling---that can create a false sense of security and utility.

\textit{Lack of dynamic, complex tasks and environments that require replanning and policy updates.}
Existing benchmarks typically cover only simple tasks with a small number of steps (e.g., 3--4) that can be planned ahead (e.g., ``read file A and put the summary in channel B''). In addition, they often fail to model dynamic environments---for example, the possibility of encountering runtime errors, as discussed in Section~\ref{sec-position-1}. Moreover, the task suite may not be diverse enough to expose the limitations of task-agnostic policies (recall the \texttt{send\_money} example in Section~\ref{sec-position-1}). We argue that benchmarks must include longer-horizon, evolving tasks and dynamic environments to assess the true utility of proposed defenses.

\textit{Lack of dynamic attack objectives.}
Most benchmark attacks focus narrowly on action hijacking to a pre-selected, static malicious tool call (e.g., instead of summarizing a webpage, the agent is coerced into posting misinformation to Slack). In these setups, the attack objective is fixed across user tasks, and the attacker-selected tool call often differs substantially from the benign one. In practice, however, attacks can be more nuanced. For example, an adversary may manipulate parameters of the \emph{same} tool call (e.g., transferring \$1000 instead of \$100), or alter the conditions under which an agent executes a plan (e.g., ``if it is not raining tomorrow, book a day hike; otherwise, purchase two movie tickets''). These task- and context-dependent objectives are harder to defend against and should be reflected in benchmark design.

\textit{Lack of payload optimization.}
Nearly all benchmark payloads are static and generated with simple heuristics. Recent work suggests that adaptive black-box attacks---for example, training a specialized attacker model against a defended system (e.g., via RL)~\cite{wen2025rl,yin2026pismith}, or using iterative optimization (e.g., genetic-algorithm-style refinement)~\cite{nasr2025attacker} to improve the payload---can be substantially more effective at surfacing prompt-injection vulnerabilities in both models and agentic systems. We therefore argue that benchmarks should support plugging in adaptive attackers, rather than relying solely on fixed attack strings.\footnote{Designing adaptive attacks for every defense can be demanding, but many effective attacks rely on an attacker LLM. At minimum, benchmarks should support adding a description of the defense algorithm and possible attack strategies to the attacker LLM's context window.}

\textbf{Question 2: What is the value of system-level defenses if they still require LLM and/or human decision making in certain cases?}
System-level defenses are often motivated by the promise of security guarantees. Yet Positions~2 and~3 imply that some security-critical decisions will still require LLM-based judgment and, in inherently ambiguous cases, human oversight. This raises a natural question: if we cannot eliminate LLMs and humans from the loop, what is the value of system-level defenses?
We argue that the key value of system-level defense research is that it provides a \emph{structured} and more \emph{explainable} way to analyze, understand, and enhance agent security. A well-designed system serves as the \emph{skeleton} on which both LLM- and human-in-the-loop components operate. %: it should define what information is observed, what decisions are made, and why those decisions are made.

\textit{(1) Structured analysis and control via secure interfaces.}
System-level design enables us to replace raw, attacker-controlled environment text with narrowly scoped artifacts and constrained choices. As discussed in Position~2, the system can ensure that the LLM judge does not consume arbitrary environment strings directly; instead, the model is invoked only for well-defined sub-judgments over structured inputs (e.g., a proposed plan/policy diff plus minimal supporting evidence). In this way, the system determines \emph{what} decision makers see and \emph{what} they are allowed to decide, significantly reducing the attack surface.

\textit{(2) More targeted and tractable LLM/human research.}
By enforcing structured inputs and bounded judgment tasks, system design makes both model-level and human-in-the-loop research more targeted and tractable. Rather than defending against arbitrary untrusted strings, model defenses can focus on specific structured sub-tasks with clearer success criteria; similarly, human-oversight interfaces can focus on decision points that are auditable and minimally burdensome. We view this kind of system-model-human co-design as a particularly promising direction.

\textit{(3) Better explainability, a bridge between rule-based and model-based security, and defense in depth.} Structured system interfaces improve explainability by making it easier to trace which evidence, rules, or judgments led to each action. Crucially, this traceability enables systematic analysis of agent failures and identification of weak points. More broadly, system-level design serves as a bridge between rule-based security enforcement and expressive model-based security-decisions, enabling stronger defense in depth without sacrificing practical utility.  
\section{Takeaways and Conclusion}\label{sec-conclusion}

In this paper, we presented our vision for system-level defenses. We articulated three core positions: the need for dynamic replanning and policy updates, the need for model-based security decisions, and the need to account for human factors in agent design. Here, we restate several key insights discussed throughout the paper. %, both explicitly and implicitly.
\begin{takeawaysbox}
\noindent \textbf{Takeaways.}

1. For general-purpose agents, many planning and policy decisions are inherently context dependent (e.g., on the ``conversation'' history). This ability to adapt decisions to evolving context is a key source of agent capability, but it also introduces significant security risk.

2. We should use LLMs \textit{only} for the parts of an agent workflow that are difficult to formalize (e.g., interpreting natural-language user requests or making complex context-dependent security judgments), and rely on programmatic components whenever possible to improve both efficiency and security.

3. When LLMs must be used for security decisions, we should tightly constrain both their inputs and their decision scope (i.e., what they can observe and what they are allowed to decide). This can significantly reduce the attack surface and enable more targeted model-level robustness research---robustness for constrained inputs and tasks, rather than for arbitrary tasks over attacker-controlled environmental text.

4. Human oversight remains unavoidable for reliable agent use, especially because ambiguous cases are unavoidable. A central challenge is reducing the burden of human intervention without sacrificing either security or utility.

\end{takeawaysbox}

Beyond our main positions, we also discussed limitations in existing agent-security benchmarks and clarified the role of system-level defenses. We argue that system-level design provides the \emph{skeleton and backbone} for securing LLM agents by connecting programmatic security rules, model-based security decisions, and human intervention. 

We hope this paper encourages further research on system--model--human co-design for building agents that are both secure and high-utility.

\section*{Acknowledgments}
We would like to thank Peiran Wang, Kexin Pei, and Joseph Lucas for their valuable feedback on the draft.
\bibliographystyle{plain}
\bibliography{ref}

@article{christodorescu2025systems,
  title={Systems Security Foundations for Agentic Computing},
  author={Christodorescu, Mihai and Fernandes, Earlence and Hooda, Ashish and Jha, Somesh and Rehberger, Johann and Shams, Khawaja},
  journal={arXiv preprint arXiv:2512.01295},
  year={2025}
}

@article{zhang2024agent,
  title={Agent security bench (asb): Formalizing and benchmarking attacks and defenses in llm-based agents},
  author={Zhang, Hanrong and Huang, Jingyuan and Mei, Kai and Yao, Yifei and Wang, Zhenting and Zhan, Chenlu and Wang, Hongwei and Zhang, Yongfeng},
  journal={arXiv preprint arXiv:2410.02644},
  year={2024}
}

@article{li2026agentdyn,
  title={AgentDyn: A Dynamic Open-Ended Benchmark for Evaluating Prompt Injection Attacks of Real-World Agent Security System},
  author={Li, Hao and Wen, Ruoyao and Shi, Shanghao and Zhang, Ning and Xiao, Chaowei},
  journal={arXiv preprint arXiv:2602.03117},
  year={2026}
}

@article{wang2026landscape,
  title={The Landscape of Prompt Injection Threats in LLM Agents: From Taxonomy to Analysis},
  author={Wang, Peiran and Li, Xinfeng and Xiang, Chong and Zhang, Jinghuai and Li, Ying and Zhang, Lixia and Wang, Xiaofeng and Tian, Yuan},
  journal={arXiv preprint arXiv:2602.10453},
  year={2026}
}

@article{ji2025taxonomy,
  title={Taxonomy, Evaluation and Exploitation of IPI-Centric LLM Agent Defense Frameworks},
  author={Ji, Zimo and Wang, Xunguang and Li, Zongjie and Ma, Pingchuan and Gao, Yudong and Wu, Daoyuan and Yan, Xincheng and Tian, Tian and Wang, Shuai},
  journal={arXiv preprint arXiv:2511.15203},
  year={2025}
}

@article{shi2025progent,
  title={Progent: Programmable privilege control for llm agents},
  author={Shi, Tianneng and He, Jingxuan and Wang, Zhun and Li, Hongwei and Wu, Linyu and Guo, Wenbo and Song, Dawn},
  journal={arXiv preprint arXiv:2504.11703},
  year={2025}
}

@article{debenedetti2025defeating,
  title={Defeating prompt injections by design},
  author={Debenedetti, Edoardo and Shumailov, Ilia and Fan, Tianqi and Hayes, Jamie and Carlini, Nicholas and Fabian, Daniel and Kern, Christoph and Shi, Chongyang and Terzis, Andreas and Tram{\`e}r, Florian},
  journal={arXiv preprint arXiv:2503.18813},
  year={2025}
}

@article{wang2025agentarmor,
  title={Agentarmor: Enforcing program analysis on agent runtime trace to defend against prompt injection},
  author={Wang, Peiran and Liu, Yang and Lu, Yunfei and Cai, Yifeng and Chen, Hongbo and Yang, Qingyou and Zhang, Jie and Hong, Jue and Wu, Ye},
  journal={arXiv preprint arXiv:2508.01249},
  year={2025}
}

@article{li2025drift,
  title={DRIFT: Dynamic Rule-Based Defense with Injection Isolation for Securing LLM Agents},
  author={Hao Li and Xiaogeng Liu and Hung-Chun Chiu and Dianqi Li and Ning Zhang and Chaowei Xiao},
  journal = {NeurIPS},
  year={2025}
}

@article{kim2025prompt,
  title={Prompt flow integrity to prevent privilege escalation in llm agents},
  author={Kim, Juhee and Choi, Woohyuk and Lee, Byoungyoung},
  journal={arXiv preprint arXiv:2503.15547},
  year={2025}
}

@article{debenedetti2024agentdojo,
  title={Agentdojo: A dynamic environment to evaluate prompt injection attacks and defenses for llm agents},
  author={Debenedetti, Edoardo and Zhang, Jie and Balunovic, Mislav and Beurer-Kellner, Luca and Fischer, Marc and Tram{\`e}r, Florian},
  journal={Advances in Neural Information Processing Systems},
  volume={37},
  pages={82895--82920},
  year={2024}
}

@article{nasr2025attacker,
  title={The attacker moves second: Stronger adaptive attacks bypass defenses against LLM jailbreaks and prompt injections},
  author={Nasr, Milad and Carlini, Nicholas and Sitawarin, Chawin and Schulhoff, Sander V and Hayes, Jamie and Ilie, Michael and Pluto, Juliette and Song, Shuang and Chaudhari, Harsh and Shumailov, Ilia and others},
  journal={arXiv preprint arXiv:2510.09023},
  year={2025}
}

@article{wen2025rl,
  title={Rl is a hammer and llms are nails: A simple reinforcement learning recipe for strong prompt injection},
  author={Wen, Yuxin and Zharmagambetov, Arman and Evtimov, Ivan and Kokhlikyan, Narine and Goldstein, Tom and Chaudhuri, Kamalika and Guo, Chuan},
  journal={arXiv preprint arXiv:2510.04885},
  year={2025}
}

@inproceedings{chen2025struq,
  title={$\{$StruQ$\}$: Defending against prompt injection with structured queries},
  author={Chen, Sizhe and Piet, Julien and Sitawarin, Chawin and Wagner, David},
  booktitle={34th USENIX Security Symposium (USENIX Security 25)},
  pages={2383--2400},
  year={2025}
}

@inproceedings{chen2025secalign,
  title={Secalign: Defending against prompt injection with preference optimization},
  author={Chen, Sizhe and Zharmagambetov, Arman and Mahloujifar, Saeed and Chaudhuri, Kamalika and Wagner, David and Guo, Chuan},
  booktitle={Proceedings of the 2025 ACM SIGSAC Conference on Computer and Communications Security},
  pages={2833--2847},
  year={2025}
}

@article{chen2025meta,
  title={Meta secalign: A secure foundation llm against prompt injection attacks},
  author={Chen, Sizhe and Zharmagambetov, Arman and Wagner, David and Guo, Chuan},
  journal={arXiv preprint arXiv:2507.02735},
  year={2025}
}

@inproceedings{hung2025attention,
  title={Attention tracker: Detecting prompt injection attacks in llms},
  author={Hung, Kuo-Han and Ko, Ching-Yun and Rawat, Ambrish and Chung, I-Hsin and Hsu, Winston H and Chen, Pin-Yu},
  booktitle={Findings of the Association for Computational Linguistics: NAACL 2025},
  pages={2309--2322},
  year={2025}
}

@article{zou2025pishield,
  title={PIShield: Detecting Prompt Injection Attacks via Intrinsic LLM Features},
  author={Zou, Wei and Liu, Yupei and Wang, Yanting and Chen, Ying and Gong, Neil and Jia, Jinyuan},
  journal={arXiv preprint arXiv:2510.14005},
  year={2025}
}

@article{hines2024defending,
  title={Defending against indirect prompt injection attacks with spotlighting},
  author={Hines, Keegan and Lopez, Gary and Hall, Matthew and Zarfati, Federico and Zunger, Yonatan and Kiciman, Emre},
  journal={arXiv preprint arXiv:2403.14720},
  year={2024}
}

@article{shi2025promptarmor,
  title={Promptarmor: Simple yet effective prompt injection defenses},
  author={Shi, Tianneng and Zhu, Kaijie and Wang, Zhun and Jia, Yuqi and Cai, Will and Liang, Weida and Wang, Haonan and Alzahrani, Hend and Lu, Joshua and Kawaguchi, Kenji and others},
  journal={arXiv preprint arXiv:2507.15219},
  year={2025}
}

@inproceedings{greshake2023not,
  title={Not what you've signed up for: Compromising real-world llm-integrated applications with indirect prompt injection},
  author={Greshake, Kai and Abdelnabi, Sahar and Mishra, Shailesh and Endres, Christoph and Holz, Thorsten and Fritz, Mario},
  booktitle={AISec@CCS},
  year={2023}
}

@inproceedings{zhan2024injecagent,
  title={Injecagent: Benchmarking indirect prompt injections in tool-integrated large language model agents},
  author={Zhan, Qiusi and Liang, Zhixiang and Ying, Zifan and Kang, Daniel},
  booktitle={Findings of the Association for Computational Linguistics: ACL 2024},
  pages={10471--10506},
  year={2024}
}

@article{wallace2024instruction,
  title={The instruction hierarchy: Training llms to prioritize privileged instructions},
  author={Wallace, Eric and Xiao, Kai and Leike, Reimar and Weng, Lilian and Heidecke, Johannes and Beutel, Alex},
  journal={arXiv preprint arXiv:2404.13208},
  year={2024}
}

@article{cutler2024cedar,
  title={Cedar: A New Language for Expressive, Fast, Safe, and Analyzable Authorization (Extended Version)},
  author={Cutler, Joseph W and Disselkoen, Craig and Eline, Aaron and He, Shaobo and Headley, Kyle and Hicks, Michael and Hietala, Kesha and Ioannidis, Eleftherios and Kastner, John and Mamat, Anwar and others},
  journal={arXiv preprint arXiv:2403.04651},
  year={2024}
}

@online{willison2023dual,
  author       = {Simon Willison},
  title        = {The Dual LLM pattern for building AI assistants that can resist prompt injection},
  year         = {2023},
  month        = apr,
  day          = {25},
  url          = {https://simonwillison.net/2023/Apr/25/dual-llm-pattern/}
}

@article{kang2025mitigating,
  title={Mitigating Indirect Prompt Injection via Instruction-Following Intent Analysis},
  author={Kang, Mintong and Xiang, Chong and Kariyappa, Sanjay and Xiao, Chaowei and Li, Bo and Suh, Edward},
  journal={arXiv preprint arXiv:2512.00966},
  year={2025}
}

@article{yin2026pismith,
  title        = {PISmith: Reinforcement Learning‑based Red Teaming for Prompt Injection Defenses},
  author       = {Chenlong Yin and Runpeng Geng and Yanting Wang and Jinyuan Jia},
  journal      = {arXiv preprint arXiv:2603.13026},
  year         = {2026},
}
\end{document}